\documentclass[%
reprint,
 amsmath,amssymb,
 aps,
prb,
]{revtex4-2}

\usepackage{graphicx}
\usepackage{dcolumn}
\usepackage{bm}
\usepackage{xcolor}

\begin{document}

\preprint{APS/123-QED}

\title{Quantum entropies of realistic states of a topological insulator }%

\author{Nicolás Legnazzi}
\email{nicolaslegnazzi@gmail.com }
\author{Omar Osenda}%
 \email{osenda@famaf.unc.edu.ar}
\affiliation{
 Facultad de Matem\'atica, Astronom\'ia, F\'{\i}sica y Comptaci\'on, 
 Universidad Nacional de C\'ordoba, and Instituto de F\'{\i}sica Enrique 
 Gaviola, CONICET, Av. Medina Allende s/n , Ciudad Universitaria, 
 CP:X5000HUA Córdoba, Argentina}

\date{\today}

\begin{abstract}
Nanowires of BiSe show topological states localized near the surface of 
the material. The topological nature of these states can be analyzed using 
well-known quantities. In this paper, we calculate the topological 
entropy suggested by Kitaev and Preskill for these states together with a new 
entropy based on a reduced density matrix that we propose as a measure 
to distinguish topological one-electron states. Our results show that 
the topological entropy is a constant independent of the parameters that 
characterize a topological state as its angular momentum, longitudinal 
wave vector, and radius of the nanowire. The new entropy is always larger 
for topological states than for normal ones,  allowing the identification 
of the topological ones. We show how the reduced density matrices 
associated with both entropies are constructed from the pure state 
using positive maps and explicitly obtaining the Krauss operators.
\end{abstract}

\maketitle


\section{\label{sec:introduction} Introduction}

Topological states appear in different materials and 
geometries \cite{Qi2011}. Their appearance changes the conducting 
properties of the material, giving place to many distinctive phenomena 
like the quantum spin  Hall effect \cite{Kane2005,Bernevig2006,Murakami2011}, 
topological superconductors \cite{Sato2017,Kezilebieke2020}, etc. The 
compound $\mbox{Bi}_2\mbox{Se}_3$ is a 3D topological 
insulant \cite{Qi2011,Liu2010,Zhang2009,Fu2007}. 

The calculation of the spectrum, spin currents, and density of 
states (DOS) of BiSe cylinders or quantum dots is the subject of 
numerous works \cite{Lou2011,Governale2020,Iorio2016,Linder2009,Gioia2019}. 
Besides first principle calculations, the kp model is a preferred tool to 
calculate the spectrum and eigenstates necessary to obtain the spin currents 
and the DOS \cite{Liu2010}. Once the eigenstates are available, it is possible 
to characterize the topological states using quantum entropies to study 
different physical regimes. 

The cylindrical geometry is well-suited to calculating the Kitaev-Preskill 
topological entropy \cite{Kitaev2006}. This quantity should be a constant, 
independent of all the quantum numbers that characterize a given topological 
state. It can only depend on the topology of the problem. But what are the 
values of this entropy for the normal in-band eigenstates that are 
eigenfunctions of the kp model? In a situation without a boundary, the 
topological entropy of a normal state should be null, but a cylindrical 
nanowire necessarily has one. So, the topological entropy for a normal 
eigenstate obtained using the kp method could be non-null, but its value 
should depend on its quantum numbers. 

Distinguishing topological from normal states can be done using different 
quantum entropies and related quantities, such as the entanglement 
spectrum \cite{Li2008, Fidkowsk2010, Qi2012, Calabrese2008, Yao2010}. To this 
end, we propose a particular reduced-density matrix whose entropy distinguishes 
the one-electron topological states of a $\mbox{Bi}_2\mbox{Se}_3 $ cylinder 
from the non-topological ones. Our study differs from those that employ 
entanglement entropy to study many-electron wave functions. The entanglement 
entropy detects the non-local character of topological states when applied, for instance,  to quantum states that are good approximations of the many-electron ground-state wave function of the fractional Quantum Hall effect, that is,  the Laughlin states \cite{Haque2007, Li2008, Zozulya2007}.

For a given eigenstate of the kp Hamiltonian, which we calculate as a superposition of a basis set functions as is usual in the Rayleigh-Ritz variational method, the proposed reduced density matrix depends on the coefficients of the superposition and integrals of the basis set functions. The Rayleigh-Ritz variational method accurately provides the band structure near the gap between the conduction and valence bands in semiconductor nanostructures when applied to kp Hamiltonians. The application of the Rayleigh-Ritz variational method to kp Hamiltonians allowed the study of electronic and optical properties of core-shell nanowires \cite{Kishore2012,Kishore2014}, edge states with and without an external magnetic field applied 
to a quantum well 
\cite{Krishtopenko2018,Krishtopenko2016,Skolanski2018,Chen2019}, the transition between resonance and bound states in quantum dots embedded in nanowires \cite{Giovenale2022a}, the entanglement entropy of edge states in quantum wells \cite{Giovenale2022b},  spin currents in topological insulators \cite{Lou2011},  amongst other physical phenomena. 

The paper is organized as follows. In Section 2, we present the kp Hamiltonian, whose eigenvalues give the band structure of a cylindrical nanowire made of $\mbox{Bi}_2\mbox{Se}_3 $ and briefly describe how to obtain a numerical approximation to the spectrum and eigenvectors using the Rayleigh-Ritz method. Section 3 deals with the calculus of the topological entropy of an approximate variational eigenvector. The topological entropy depends on the von Neumann entropy of several reduced-density matrices. Obtaining each one of these matrices implies tracing out a real-space partition from a pure density one \cite{Kitaev2006, Sterdyniak2012}.  

We present in Section IV the mode-dependent reduced density matrix (RDM), $\rho_{MD}$, which contains information about a variational state through the coefficients of the variational expansion and spatial integrals of the basis set functions. We show how the von Neumann entropy and the entanglement spectrum \cite{Li2008} of the mode-dependent RDM allow us to distinguish between non-topological and topological states. 

In Section V, we use quantum state processes \cite{Sudarshan1961,Nielsen2000} as an alternative way to construct the mode-dependent RDM and some of the RDM necessary to calculate the topological entropy. Using the quantum process tomography algorithm described in Reference \cite{Ahmed2022}, the quantum process results in a sum of Kraus operators determined by a gradient-descent algorithm. The quantum process numerically calculated predicts an RDM, $\tilde{\rho}$, slightly different from the one intended, $\rho'$. We compare both matrices, calculating their fidelity. Finally, in Section IV, we summarize and discuss our results.

\section{\label{sec:model} Model and Hamiltonian} 

We consider a cylindrical nanowire made of $Bi_2 Se_3$, with a constant radius on the tens of nanometers and infinitely long in the axial direction. 

To obtain the band structure and eigenstates, we employ the $\mathbf{k}\cdot\mathbf{p}$ Hamiltonian given by
\begin{equation}\label{eq:Hamiltonian}
H_{0}=\varepsilon(\mathbf{k})+\left(\begin{array}{c c c c}{{M(\bf{k})}}&{{B(k_{z})k_{z}}}&{{0}}&{{A\left(k_{\parallel}\right)k_{-}}}\\ {{B(k_{z})k_{z}}}&{{-M(\bf{k})}}&{{A(k_{\parallel})k_{-}}}&{{0}}\\{{0}}&{{A\left(k_{\parallel}\right)k_{+}}}&{{M(\bf{k})}}&{{-B(k_{z})k_{z}}}\\ {{A\left(k_{\parallel}\right)k_{+}}}&{{0}}&{{-B(k_{z})k_{z}}}&{{-M(k_{z})k_{z}}}\end{array}\right),
\end{equation}
where
\begin{itemize}
    \item $\varepsilon(\mathbf{k})=C_{0}+C_{1}k_{z}^{2}+C_{2}k_{\parallel}^{2}$,
    \item $M(\mathbf{k})=M_{0}+M_{1}k_{z}^{2}+M_{2}k_{\parallel}^{2}$,
    \item $B(k_{z})=B_{0}$,
    \item $A(k_{\parallel})=A_{0}$.
\end{itemize}
The Hamiltonian in Equation~\ref{eq:Hamiltonian} was introduced by Zhang \cite{Zhang2009} and collaborators to adjust the band structure found for different materials showing topological states. We consign the parameters that define the Hamiltonian in Equation~\ref{eq:Hamiltonian}  in  Table~\ref{tab:parametros}.
\begin{table}[h]
\begin{center}
\begin{tabular}{| c | c | c |}
\hline
Parameter & Units &  \\ \hline
$C_{0}$ & $eV$ & -0.0068\\
$M_{0}$ & $eV$ & 0.28\\
$C_{1}$ &  $eV\cdot A$ & 1.3\\
$M_{1}$ & $eV\cdot A$ & -10.0\\ 
$C_{2}$ & $eV\cdot A^{2}$ & 19.6 \\
$M_{2}$ & $eV\cdot A^{2}$ & -56.6 \\
$A_{0}$ & $eV\cdot A^{2}$ & 4.1 \\
$B_{0}$ & $eV\cdot A$ &  2.2 \\ \hline
\end{tabular}
\caption{The Table contains the values of the parameters that enter into the four band $\mathbf{k}\cdot\mathbf{p}$ Hamiltonian of a $\mbox{Bi}_2 \mbox{Se}_3$ nanowire. }
\end{center}
\label{tab:parametros}
\end{table}

We calculate approximate eigenvalues and eigenfunctions using the Rayleigh-Ritz variational method, which reduces the eigenvalue problem
\begin{equation}
H f = e f, 
\end{equation}
where $H$is the $\mathbf{k}\cdot\mathbf{p}$ Hamiltonian, to an algebraic one. The  Rayleigh-Ritz method is suitable for calculating the band structure of nano-structures near the gap between the conduction and valence bands and for states lying inside it, for instance, to find the topological states in three-dimensional topological insulators, to study the transition between localized and resonance states in quantum dots, to study properties of states in quantum wells in the Quantum Spin Hall Effect regime, etc. We consign the details about the implementation of the method to Appendix \ref{ap:variational-method}.

\section{The topological entropy}

The study of topological states has led to different entropic-like 
quantities as, for instance, the real-space entropy, the topological 
entropy of Kitaev-Preskill, and so on. Also, it is worth mentioning 
the entanglement spectrum or, in the case of studies dealing with 
topological states in spin chains, the Renyi entropies. 

\begin{figure}[h]
\includegraphics[width=7cm]{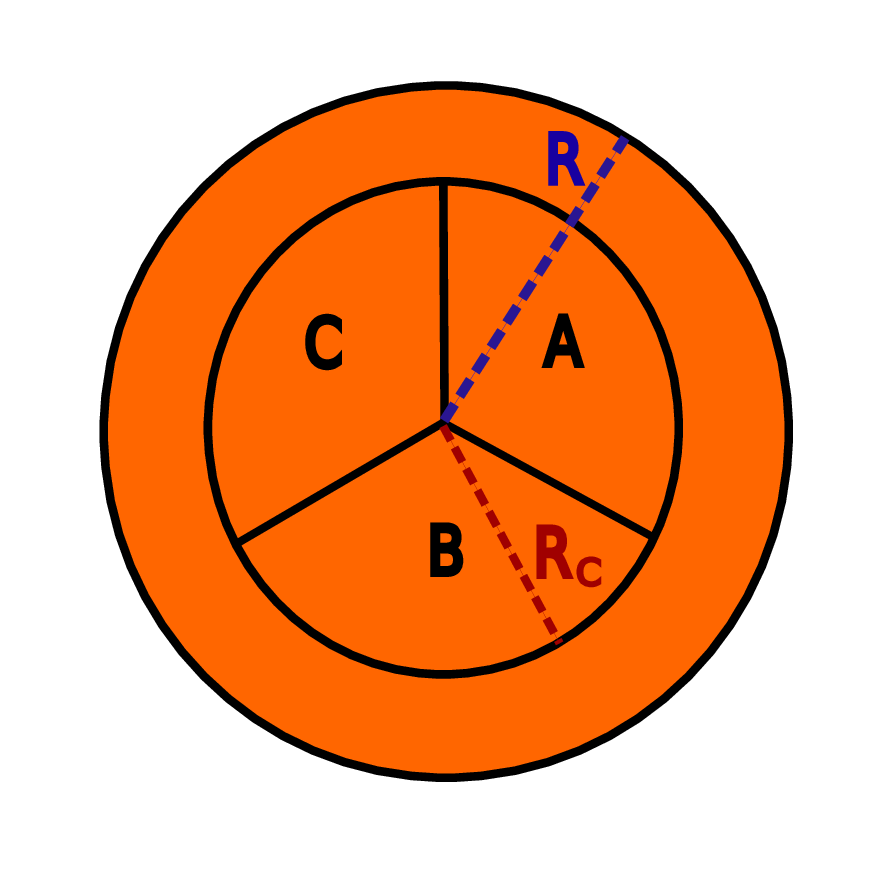}
\centering
\caption{The cartoon depicts the main geometric elements that enter 
into the calculus of the topological entropy and other quantities in 
Sections 2, 3, 4, and 5. $R$ is the radius of the nanowire, $R_c$ is the 
radius associated to the circular sectors $A$, $B$, and $C$. The central 
angle of each sector is equal to $2\pi/3$. The blue dashed line 
corresponds to $R$ while the red one corresponds to $R_c$. }
 \label{fig:regioneskitaev}
\end{figure}

Tracing out different subspaces of the whole Hilbert space leads to one 
or other entropy. For instance, for a given multiparticle wavefunction $\psi$, the real-space entropy
\begin{equation}
S_A(\rho_A) = - \mbox{Tr}(\rho_A \log{(\rho_A)}),
\end{equation}
is calculated from the reduced density operator defined by
\begin{equation}
\rho_A = \frac{1}{Z} \int_{\bar{A}}  \, \psi^{\star}(\vec{x})   \psi(\vec{x})  d\vec{x}^{\prime},
\end{equation}
where $\bar{A}$ is the spatial region outside region $A$, $Z$ is a normalization constant, and the integral involves a subset $\vec{x}^{\prime}$ of all the particle coordinates.  

From the definition of the reduced density operator above, it is clear that the value of the real-space entropy could depend on the size of the spatial region. To avoid this problem, Kitaev and Preskill proposed to trace out over a set of spatial sectors and combine the corresponding real-space entropies to single out the {\em topological entropy} of the quantum state.

Following the argument in Reference \cite{Kitaev2006}, we consider three regions, $A$, $B$, and $C$, as shown in the cartoon in Figure~\ref{fig:regioneskitaev}. The triangular sectors are defined by
\begin{align}
&A : \varphi \in  \left[-\pi/6,\pi/2\right] \quad \mbox{and} \quad \rho \leq R_C, \\
&B: \varphi \in  \left[7\pi/6,11\pi/6\right] \quad \mbox{and} \quad \rho \leq R_C, \\
&C: \varphi \in  \left[\pi/2,7\pi/6\right] \quad \mbox{and} \quad \rho \leq R_C .
\end{align}
Tracing out the spatial region {\em outside} $A$, $B$, $C$, or combinations of them, the topological entropy is given by
\begin{equation}\label{eq:topo-entropy}
S_t = S_A  + S_B +S_C - S_{AB} - S_{BC} - S_{AC} + S_{ABC} .
\end{equation}

In our case, we calculate $\rho_A$ as follows
\begin{equation}
\rho_A = \int_D |\psi(\rho,\varphi,z)\rangle \langle \psi(\rho,\varphi,z)| \; \rho \, d\rho \, d\varphi,
\end{equation}
where $|\psi\rangle$ is one of the eigenvectors obtained using the Rayleigh-Ritz method, and the integral is over the surface of the disk of radius $R$ minus the triangular sector $A$, see Figure~\ref{fig:regioneskitaev}. Proceeding in this way, $\rho_A$, $\rho_B$, and so on, are given by $4\times4$ matrices. The corresponding von Neumann entropies are given by
\begin{equation}
S_A(\rho_A) = - \sum_i \lambda_i \log(\lambda_i),
\end{equation}
where the $\lambda_i$ are the eigenvalues of $\rho_A$.

The variational eigenvectors $|\psi\rangle$ are complex column vectors that depend on the three cylindrical coordinates and are labeled by two quantum numbers, $k_z$ and $L$,
\begin{widetext}
\begin{equation}\label{eq:variational-basis}
|\psi_{L,k_{z}}\rangle=\sum_{n}^N \left(\!\!\!\begin{array}{c}{{b_{L,k_{z},n,\uparrow}A_{L,n}J_{L}\left(\alpha_{n}^{L}\rho /R\right)e^{i L\varphi}}}\\ {{c_{L,k_{z},n,\uparrow}A_{L,n}J_{L}\left(\alpha_{n}^{L}\rho /R \right)e^{i L\varphi}}}\\ {{b_{L,k_{z,n,\downarrow}}A_{L+1,n}J_{L+1,n}\left(\alpha_{n}^{L+1}\rho/R\right)e^{i(L+1)\varphi}}}\\ {{c_{L,k_{z,n,n,\downarrow}}A_{L+1,n}J_{L+1,n}\left(\alpha_{n}^{L+1}\rho/R\right)e^{i(L+1)\varphi}}}\end{array}\!\!\!\right)e^{i k_{z}.z},
\end{equation}
\end{widetext}
where  $c_{L,k_{z},n,\uparrow}$, $c_{L,k_{z},n,n,\downarrow}$, $b_{L,k_{z},n,\uparrow}$ and 
$ b_{L,k_{z},n,\downarrow}$ are the linear variational coefficients,  $J_L$ is the Bessel function with index $L$, the $\alpha_n^L$ are its roots, and $A_{L,n}$ is a normalization constant.  Note that the number of roots employed coincides with the number of basis set functions used. 

Lou and collaborators \cite{Lou2011} used the variational basis set in Equation \eqref{eq:variational-basis} to calculate the band structure of $\mbox{Si}_2 \mbox{Be}_3$ nanowires. They focused on nanowires with a diameter of  120 nanometers, and for these nanowires, they showed that there are two topological states for each pair $(k_z, L)$. Of the two topological states, one has energy closer to the conduction band, while the other lies closer to the valence band. We will refer to the former state as the upper energy state and to the latter as the lower energy state, respectively. We will also focus on nanowires with a diameter of 120 nanometers for comparison purposes. 

\begin{figure}[h]
\includegraphics[width=7cm]{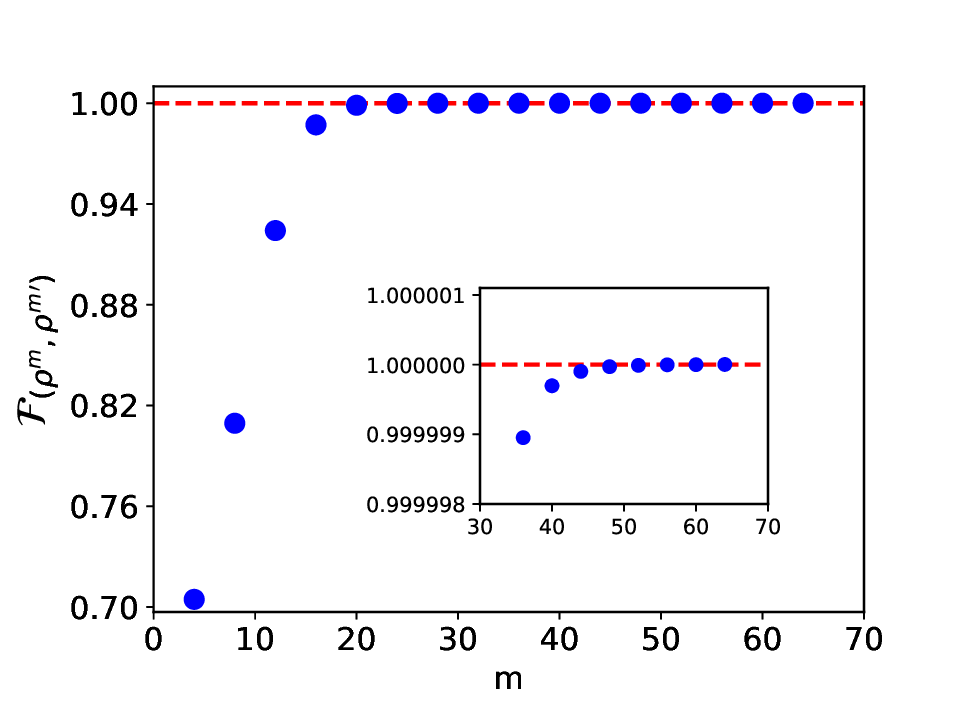}
\centering
\caption{The fidelity between two RDM, $\mathcal{F}(\rho^m,\rho^{m'})$ vs the 
basis set size $m$. The values of $m$ and $m'$ are picked as successive 
values of the set ${20,24, ..60,64}$, with $m>m'$. The solid blue dots 
correspond to the calculated values, while the black line is a guide and 
corresponds to the unity value.The reduced density matrix is 
the $\rho_{ABC}$ obtained for the topological state with 
$L=0$, $k_z=0.1$, and larger energy.}
 \label{fig:convergence-rho}
\end{figure}

As is usual with states or eigenvalues calculated using the Rayleigh-Ritz 
variational method, the larger the number of the basis set functions, the 
greater the accuracy.  If $\rho^{m}$ is an RDM obtained from a variational 
eigenvector with $m$ basis set functions, we assess the accuracy of the 
variational method by comparing two RDM with different values of $m$. The 
fidelity between a succession of $\rho_A$ matrices obtained for one of the 
two topological states with $L=0$, each matrix calculated with a different 
number of basis set functions, is shown in Figure~\ref{fig:convergence-rho}.  

Note that, while each matrix on the succession is normalized, $\mbox{Tr}(\rho)=1$, the fidelity approaches the unity for larger basis sets. This convergence indicates that the successive matrices are more and more similar. Consequently, in what follows, we will present results obtained with the basis set with $N=40$. The convergence of the energy values of the topological states is even better. For given values of $k_z$ and $L$, the figures obtained using the two larger basis sets have a relative difference of less than $1\times 10^{-4}$.

\begin{figure}[h]
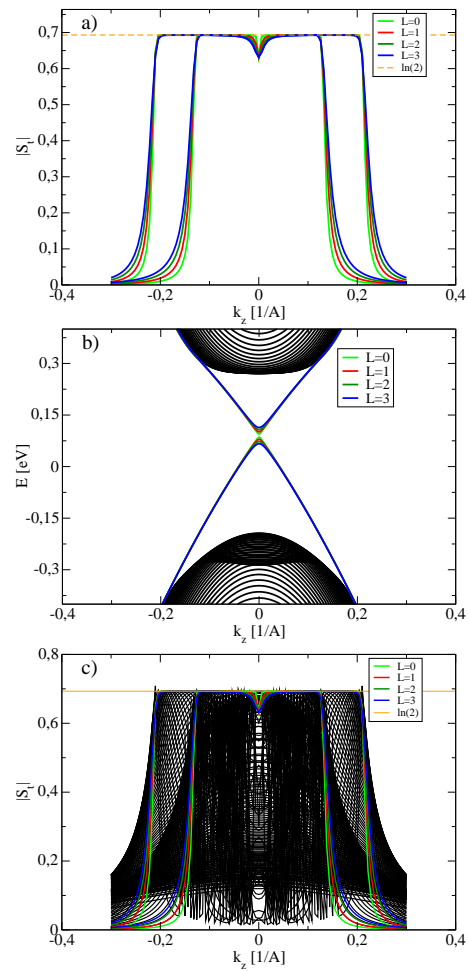

\includegraphics[width=6cm]{fig-3a.eps}
\includegraphics[width=6cm]{fig-3b.eps}
\includegraphics[width=6cm]{fig-3c.eps}
\centering
\caption{The panels show the topological entropy and eigenvalues obtained using the variational method. a) the panel shows the modulus of the topological entropy $S_t$, Equation  8,  of topological states with quantum numbers $L=0,1,2$ and $3$ versus the wave number $k_z$. The dashed line corresponds to $\gamma=\log{2}$. It is clear that $|S_t|\sim \log{2}$  for an ample range of $k_z$ values. b) The variational spectrum near the gap. The black curves correspond to the valence and conduction bands, while the colored ones correspond to the topological states. Accordingly, these energies lie inside the gap. For each value of $k_z$ and $L$, there are two topological states, one with an eigenvalue nearer to the conduction band (the upper energy value) and the other with its eigenvalue closer to the valence band (the lower energy value). Comparing both panels, it is clear that the sudden drop in the value of the topological entropy occurs when the energy of the corresponding state is close enough to the bands. c) The modulus of the topological entropy calculated for all the eigenstates whose eigenvalue we plotted in b).}
 \label{fig:constant-topo-entro}
\end{figure}

The topological entropy in Equation \eqref{eq:topo-entropy} should be independent of the quantum numbers that label the topological quantum state. Figure~\ref{fig:constant-topo-entro}a shows this property, where we plot the topological entropy of all the variational eigenstates obtained for different values of $k_z$ and $L$ as a function of the wave number $k_z$. The colored lines correspond to topological states with $L=0,1,2,3$. The topological entropy of the different topological states is constant as a function of $k_z$, except when its energy is close enough to a band, in which case the entropy value drops to zero. 

In panel b) we show the behavior of the variational eigenvalues near the gap. The color code used to depict the energy of the topological states is the same one used in panel a) for the corresponding topological entropies. The black curves correspond to the valence and conduction bands and contain all the other eigenvalues obtained from the variational method. These are the eigenvalues of the normal non-topological states.  

Comparing both panels, a) and b), it is clear that the topological entropy drops to zero when the energy of the topological state is close to the conduction or the valence bands. 

The topological states reduced density matrices are difficult to calculate accurately close to $k_z=0$. Because of this, the topological entropy curves show a little slump near $k_z=0$. The difficulty increases for larger values of $L$ and, consequently, the deeper one corresponds to the larger  $L$. 

Finally, in Figure 3c, we show the topological entropy calculated for all the states whose eigenvalues we plotted in panel b). Again, the colored curves correspond to the topological states, while the black ones correspond to the normal non-topological ones. There are a few salient traits worth commenting on. The topological entropy of the non-topological states depends on $k_z$, and eventually,  its value reaches up to $\log{2}$. Since we are dealing with a finite system, the length scale associated with the different eigenstates is finite and comparable with $R_c$, except for the topological states. So, changing the value of $R_c$ would change the topological entropy of a non-topological state but not the topological entropy of a topological one.

\begin{figure}[ht]
\includegraphics[width=7cm]{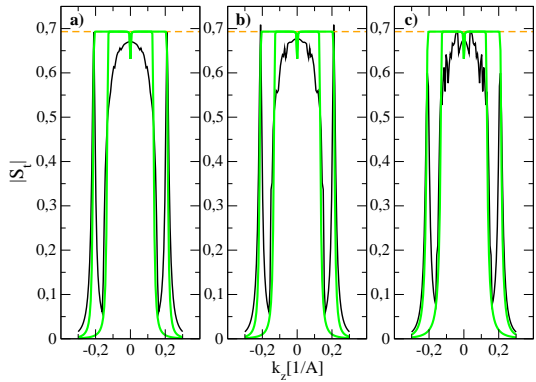}
\centering
\caption{The Figure shows the behavior of the modulus of the topological entropy for the two topological states with quantum number $L=0$  as a function of the wave number $k_z$. The green curves in each panel correspond to the topological entropy calculated with a particular $R_c$ value. a) $R_c=R/3$, b) $R_c=R/2$, and c) $R_c=2R/3$. The black curve in each panel depicts the behavior of the topological entropy calculated for a non-topological state $\psi^n(k_z, L)$, where $n$ indicates that the state has an eigenenergy $E_n$, with $n$ fixed and such that the eigenenergy lies well inside the valence/conduction band. It is easy to appreciate that the topological entropy of the topological states remains constant, irrespective of $R_c$. On the other hand, the entropy corresponding to the non-topological states changes its value when $R_c$ changes. .}
 \label{fig:entro-topo-tres-radios}
\end{figure}

The topological entropy should also be independent of the length of the circular curve that defines the triangular sectors  $A$, $B$, and $C$ or, equivalently, of the radius $R_c$. Figure~\ref{fig:entro-topo-tres-radios} shows the topological entropy as a function of the wave number for three different values of $R_c$. As the Figure clearly shows, $S_t$  is a constant with very high precision.

\section{Mode-dependent reduced density matrix, its entropy and  entanglement spectrum}
\label{sec:mode-entropy}

As the results in the previous Section show, the topological entropy is effectively a constant for the topological states up to the numerical precision. Nevertheless, this entropy is non-null for the non-topological ones. Its value depends on both quantum numbers $k_z$,  $L$, and the radius $R_c$. Note that $\log(2)$ bounds from above the values that achieve $|S_t(\rho_{nt})|$, where $\rho_{nt}$ is the density matrix of  a non-topological state. For some values of $k_z$, $L$ and $R_c$ $|S_t(\rho_{nt})|=log(2)$, which renders the Kitaev-Preskill entropy useless in this context to distinguish a topological state from normal ones. 

In this Section, we propose a reduced-density matrix whose entropy for topological states is larger than the entropy of the normal ones. The reduced-density matrix depends on the variational coefficients and spatial integrals of the variational basis set functions. 

Writing the variational eigenfunctions as 
\begin{equation}
|\psi\rangle = \sum_{n,j} c(j, L,n) f_{j,L,n}(\rho,\varphi) e^{ik_z z}|j\rangle,
\end{equation}
we define the mode-dependent reduced density matrix $\rho_{MD}$, with matrix elements
\begin{align}
\left[\rho_{MD}\right]_{j, L,n; i,L^{\prime},m} = &\frac{1}{\mathcal{N}}c(j,L,n) (c(i,L^{\prime},m))^* \times \nonumber \\ \label{eq:rho-MD}
&\int_{\Omega} f_{j,L,n}(\rho,\varphi) (f_{i,L^{\prime},m})^*(\rho,\varphi) \, \rho \, d\rho\,d\varphi ,
\end{align}
where $\mathcal{N}$ is a normalization constant such that
$\mbox{Tr}(\rho_{MD})=1$, $\Omega$ is the ring outside $ABC$, and  $\rho_{MD}$ is a $4N \times 4N$ matrix. 

The mode-dependent RDM is closely related to the pure state given by
\begin{equation}\label{eq:rho-pure}
\rho_p = |\psi\rangle \langle\psi|,
\end{equation}
which has matrix elements
\begin{equation}
\rho_p = c(j, l,n) (c(i,l^{\prime},m))^*,
\end{equation}
in the variational basis and is also a $4N\times 4N$ matrix.  

In the next Section, we show that $\rho_{MD}$ is also given by
\begin{equation}
\rho_{MD}= \sum_j K_j \rho_p K^{\dagger}_j,
\end{equation}
{\em i.e.}, $\rho_{MD}$  results from applying a  quantum process to $\rho_p$, but before we want to analyze its spectrum and the behavior of the von Neumann entropy of $\rho_{MD}$ as a function of $k_z$. 

\begin{figure}[ht]
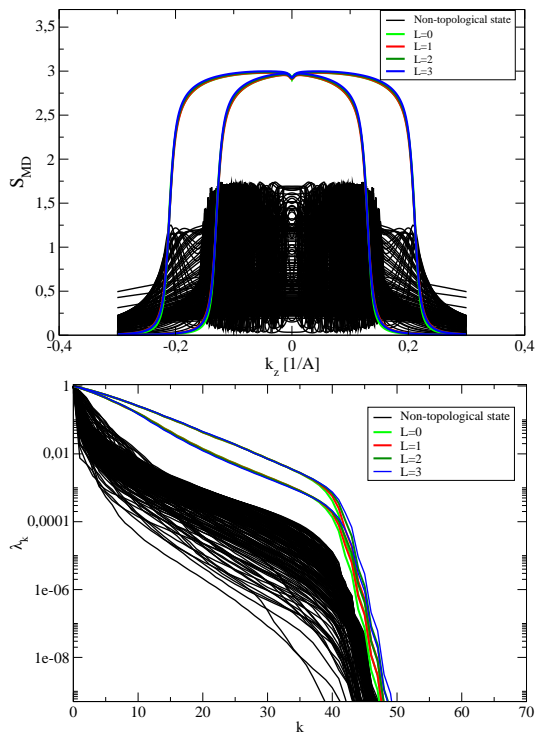

\includegraphics[width=7cm]{fig-5a.eps}
\includegraphics[width=7cm]{fig-5b.eps}
\centering
\caption{The Figure shows the mode-dependent quantum entropy and the spectrum of the mode-dependent reduced density matrix. a) The mode-dependent entropy for all the eigenstates shown in Figure 3b versus the wave number $k_z$. For the topological states, depicted with colored lines, the entropy is independent of the quantum number $L$ and shows a smooth dependence with $k_z$. Again, when the energy of the topological states becomes close enough to the bands, the value of the entropy drops to zero. b) The spectrum of the mode-dependent RDM. The eigenvalues are ordered from larger to smaller. The spectrum of the mode-dependent RDM for topological states and the normal non-topological ones correspond to the green and black lines, respectively. The entropies shown in a) were calculated using these spectra. Note that the vertical scale is logarithmic, and the different behavior exhibited by both kinds of spectra. }
 \label{fig:entro-mode-dependent}
\end{figure}

Figure~\ref{fig:entro-mode-dependent} shows a) the von Neumann entropy of the mode-dependent RDM, $(S\rho_{MD})$,  and b) the spectrum of $\rho_{MD}$ as functions of $k_z$ and the eigenvalue number, respectively. The data in panel a) shows that the entropy of the topological states has larger values than the entropy of the normal ones, except when the energy of the topological states is too close to the bands. Moreover, up to the numerical accuracy, its value depends only on the quantum number $k_z$, but not depends on the angular momentum $L$.

On the other hand, Figure 5 b) shows the mode-dependent RDM spectrum obtained corresponding to all the variational eigenvectors calculated with $N=70$. Note that the vertical scale is logarithmic. The spectra of the topological states show a decaying compatible with $\lambda_k \sim e^{-\alpha k}$ over an ample range of values, where $\alpha$ is a constant. This behavior points to the topological character of quantum states in different physical systems. Figure 6 shows the behavior of the spectrum of the mode-dependent RDM calculated using three different basis set sizes. 

\begin{figure}[h]
\includegraphics[width=7cm]{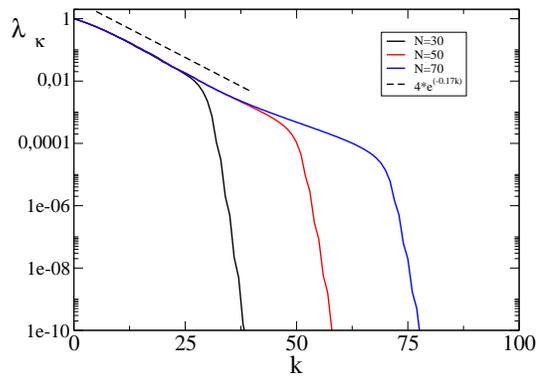}
\centering
\caption{The figure shows the spectrum of the mode-dependent reduced density matrix calculated for a topological state with quantum numbers $L=0$, and $k_z=0.1$, with three different basis set sizes $N$. The black curve corresponds to $N=30$, the red to $N=50$, and the blue to $N=70$, respectively. The black dashed line corresponds to an exponential function and is a guide to the eye. }
 \label{fig:exponential-decay}
\end{figure}
The black, red, and blue curves shown in Figure~\ref{fig:exponential-decay}  correspond to basis sets with $N=30$, $50$, and $70$ functions, respectively. The dashed black line is a guide to the eye. The number of eigenvalues that show an exponential decay grows with the number of basis functions used to obtain the RDM, although the increase is slow. This trait is another manifestation of the difficulties inherent to the obtention of the mode-dependent RDM or, more generally, how difficult it is to numerically calculate other quantities related to topological states beyond their spectrum. The entanglement spectrum, which is given by
\begin{equation}
\zeta_k = -\log(\lambda_k),
\end{equation}
together with the exponential decay of  $\lambda_k$ result in 
\begin{equation}
\zeta_k = c+ \alpha k.
\end{equation}

The abrupt decay of the eigenvalues beyond the exponential regime, shown in Figure~\ref{fig:exponential-decay}, marks when the numerical approximation becomes inaccurate, rendering all the remaining values indistinguishable from zero. 

\section{Quantum process tomography}
\label{sec:qpt}

The results in the previous Section suggest different choices for the reduced density matrices if we wish to distinguish between topological and normal states using entropic-like quantities, at least when dealing with approximate ones obtained from phenomenological Hamiltonians. 

For a given known quantum state, we could obtain the mode-dependent RDM using different basis sets, not only the particular one that we employ to implement the variational method. So, up to a point,  some features of the entropy of the mode-dependent RDM should depend on the basis set chosen. 

Instead of implementing our calculations in a different basis set, we prefer to show that the mode-dependent RDM is equivalent to many others by showing how to obtain it from the pure state in Equations 13 and 14 using quantum processes. Doing this has the twofold purpose of not dealing with the complicated calculations inherent to a change of functions basis set and showing that RDM obtained through operational processes leads to entropies that distinguish topological from normal states.

A quantum process takes a given RDM $\rho$ to another one, $\rho^{\prime}$ as follows
\begin{equation}\label{eq:quantum-process-map}
\rho^{\prime} = \mathcal{E}(\rho) .
\end{equation}
The quantum process $\mathcal{E}$ is a linear superoperator, and $\rho$ and $\rho^{\prime}$ as operators do not necessarily act on Hilbert spaces of the same dimension. The particular $\mathcal{E}$ that relates a given pair of $\rho$ and $\rho^{\prime}$ is determined using quantum process tomography, which usually is an expensive calculation. 

There are numerous algorithms to calculate $\mathcal{E}$ \cite{Ahmed2022,Fiurasek2001,Sacchi2001,Anis2012,Schultz2019,Knee2019,
Surawy2022,Baldwin2014,Teo2020,Xue2022}, which depend on the representation used for the process and the dimensionality of the involved RDM. We use the method proposed by Ahmed {\it et al.} \cite{Ahmed2022}, which assumes a Kraus representation for the quantum process
\begin{equation}\label{eq:kraus-representation}
\mathcal{E} = \sum_j K_j \rho K_j^{\dagger},
\end{equation}
and that the dimensionality of both Hilbert spaces, where $\rho$ and $\rho^{\prime}$ act, is a power of two. Ahmed {\em et al.} define a cost function that depends on the two RDM, the Kraus operators, $K_j$, and learn them using a gradient-descent method. 

Starting with randomly chosen initial Kraus operators, the non-negative cost function $\mathcal{L}$ is minimized up to values near zero to learn the optimal Kraus operators. See Appendix B for details about the calculation. We studied two cases. For the first case, $\rho$ is the pure state in Equation~\eqref{eq:rho-pure}, and $\rho^{\prime}=\rho_{ABC}$. For the second one, $\rho$ is again the pure state in Equation~\eqref{eq:rho-pure}, and $\rho^{\prime}$ is the mode-dependent RDM. 

The cost function is given by
\begin{equation}
\mathcal{L}(\mathbb{K})= \sum_j \left[d_{j} - \mbox{Tr} \left[ \mathcal{M}_j \left( \sum_l K_l \rho_p K_l^{\dagger} \right)\right] \right]^2 + \lambda ||\mathbb{K}||_1, 
\end{equation}
where 
\begin{equation}
d_{j}= \mbox{Tr} (\mathcal{M}_j \rho^{\prime}),
\end{equation}
$\mathcal{M}_j$ is a set of measurements, and $\mathbb{K} = \left[K_1,K_2, \ldots,K_{n_k}\right]$ is a $n_k M\times M'$ matrix formed with all the $M\times M'$ Kraus operators . The matrix norm is given by
\begin{equation}
||A||_1 = max_j \sum_i |A_{ij}| ,
\end{equation}
and $\lambda\geq 0$ is the strength of the regularization imposed on $\mathcal{L}$. $\lambda$ is a hyperparameter of the minimization process and can be fixed or optimized. The Kraus operators must fulfill the condition
\begin{equation}
\sum_l K_l^{\dagger} K_l = \mathbb{I} .  
\end{equation}

As we said previously, we consider two cases $\rho' =\rho_{ABC}$ and $\rho'=\rho_{MD}$.  In the following, we focus on the former case and will return to the latter near the end of the Section.

\subsection*{ $\rho'=\rho_{ABC}$ case}

In Reference~\cite{Ahmed2022}, Ahmed et al. discussed all the necessary details to implement the minimization of $\mathcal{L}$ and provided the code to reproduce their results. To obtain the results found in this Section, we adapted the code. For the case where $\rho'=\rho_{ABC}$, the only adaptation needed arises from the different dimensions of the  Hibert space where $\rho$ and $\rho'$ act. 

When the dimension of both Hilbert spaces is the same, it is reasonable to use random unitary square matrices as the initial random Kraus operators needed by the minimization algorithm. So,
\begin{equation}
\frac{1}{n_k} \sum U_l U_l^{\dagger} =1,
\end{equation}
results in
\begin{equation}
K_l = \frac{1}{\sqrt{n_k}} U_l .
\end{equation}
There are different numerical methods to construct square random unitary matrices. 

\begin{figure}[h]
\includegraphics[width=7cm]{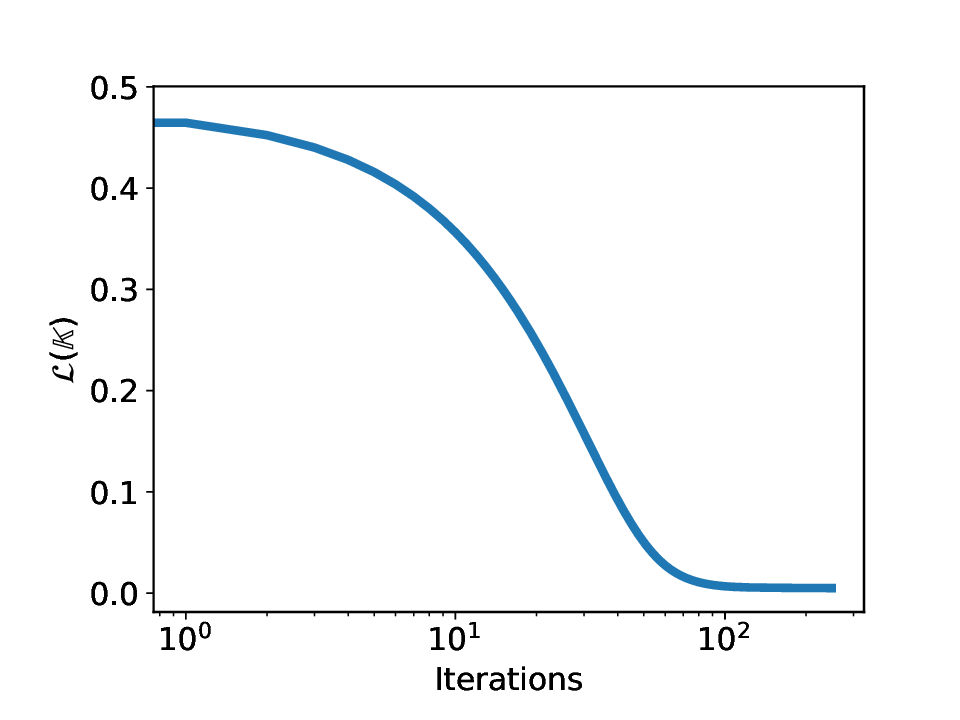}
\centering
\caption{The figure shows the typical behavior of the cost function $\mathcal{L}$ versus the number of iterations of the gradient-descent method. As the algorithm requires,  we instantiate randomly chosen initial Kraus operators into Equation 21 to produce the initial value of the cost function. The data corresponds to the case where the target  $\rho_{ABC}$ is the RDM of the topological state with $k_z=0.1$, $L=0$, and lower energy see Figure 3. After a few hundred iterations, the cost function value becomes smaller than the tolerance value, and the algorithm halts. In this case, the algorithm employs $n_k=20$ Kraus operators.}
 \label{fig:loss-function-kitaev}
\end{figure}

For the case when $\rho'=\rho_{ABC}$ is a $4\times 4$ matrix and $\rho_p$ is an $M\times M$matrix, with $M$ a multiple of four, the quantum process $\mathcal{E}$ requires $n_k$ $4\times M$ Kraus operators. Each Kraus operator is composed of $M/4$ blocks. We choose, as the initial Kraus operators, $4\times M$ matrices such that they have only one $4\times 4$ block different from zero, and this block is a random unitary matrix. If $K_1$ has only the first block different from zero, $K_2$ the second one, and so on, it is clear that the condition $Tr =1$ requires a renormalization of each random unitary matrix by a factor equal to $\frac{1}{\sqrt{n_k}}$.  

The measurements $M_j$ are given in terms of the eigenvectors of all the operators of the form $\sigma^{\beta}\otimes\sigma^{\gamma}$, where $\beta=x,y,z$ and also $\gamma=x,y,z$ that is,  all the operators that are the tensorial product of two Pauli matrices. 

Figure~\ref{fig:loss-function-kitaev} shows the typical behavior of the cost function as a function of the number of iterations performed by the gradient-descent minimization method. The algorithm that implements the gradient-descent minimization method has a tolerance parameter. Once the cost function value becomes smaller than the tolerance value,  the algorithm does not further iterate. We used a tolerance value of $0.01$ for $n_k=64$. Imposing lower values for the tolerance does not necessarily improve the results obtained and, in some cases, leads to oscillations in the cost function behavior. Since the cost function value does not become zero, the predicted RDM will differ from the one employed as the target RDM, that is, 
$\tilde{\rho}_{ABC} =\tilde{\mathcal{E}}(\rho_p)\neq \rho_{ABC}$, where $\tilde{\mathcal{E}}$ is the quantum process found using a finite non-null tolerance. Other sources of errors that prevent the predicted RDM from becoming equal to the target RDM are the number of Kraus operators used, numerical precision, etc. 

\begin{figure}[h]
\includegraphics[width=7cm]{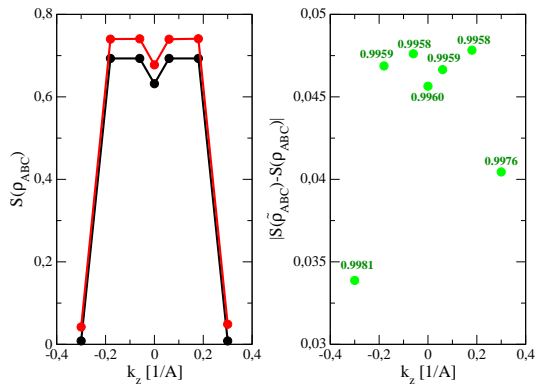}
\centering
\caption{The von Neumann entropy for both the target reduced density matrix $\rho_{ABC}$ and the predicted one $\tilde{\rho}_{ABC}$ calculated for several topological states, all with $L=0$ and different wave number $k_z$. a) The von Neumann entropy $S(\rho_{ABC})$ is shown using black solid dots, while $S(\tilde{\rho}_{ABC})$ is shown using red solid dots. Both entropies show a plateau for $k_z$ values consistent with the one observed in Figure 3. Note that the topological states used correspond to the upper branch of energies. b) This panel shows  $|S(\tilde{\rho}_{ABC}) - S(\rho_{ABC})$ using green solid dots, versus $k_z$. We calculated the modulus of the difference between the entropies with the data contained in panel a). The number next to each dot is the fidelity between both sets of matrices,  $\mathcal{F}(\rho_{ABC}(k_z, L),\tilde{\rho}_{ABC}(k_z, L))$. }
 \label{fig:predicted-kitaev}
\end{figure}

Figure \ref{fig:predicted-kitaev} shows a comparison between the values obtained for the $S(\rho_{ABC})$ entropy and the entropy calculated with the RDM resulting from the quantum process $\tilde{\rho}_{ABC}$. In panel a), the entropy $S(\rho_{ABC})$ is shown using black solid dots, while for $S(\tilde{\rho}_{ABC})$, we use red solid dots.

 It is easy to appreciate that the quantum process consistently results in RDM that gives larger entropy values than those corresponding to RDM calculated with the variational eigenvectors, as in  Eq. 9. Panel b) shows the modulus of the difference between $S(\rho_{ABC})$  and $S(\tilde{\rho}_{ABC})$ using green points. The differences shown correspond to the data shown in panel a). The number next to each green dot is the fidelity $\mathcal{F}(\rho_{ABC},\tilde{\rho}_{ABC})$. Despite the excellent fidelity between both sets of reduced-density matrices, the calculated, $\rho_{ABC}$, and the predicted by the quantum process, $\tilde{\rho}_{ABC}$, the differences in their corresponding entropies can be as large as $0.047$ or a relative error of 7$\%$. Nevertheless, note that the predicted entropy is also a constant where the calculated entropy is a constant and that its value also drops to zero for the values of $k_z$ where the energy of the topological state becomes close enough to the band, see panel a).

\subsection*{$\rho'= \rho_{MD}^{\prime}$ case}

In this case, the quantum process takes $\rho_p$, which is  $4M\times 4M$ matrix, to a predicted matrix  $\rho_{MD}^{\prime}$, which is also a $4M\times 4M$ matrix. In our case, $M=64$ for the largest basis set size, resulting in  $2^8\times 2^8$ matrices. 

\begin{figure}[h]
\includegraphics[width=7cm]{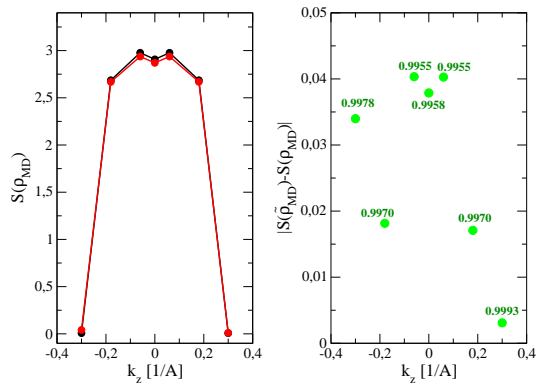}
\centering
\caption{The von Neumann entropy for both the target mode-dependent reduced density matrix $\rho_{MD}$ and the predicted one $\tilde{\rho}_{MD}$ calculated for several topological states, all with $L=0$ and different wave number $k_z$. a) The von Neumann entropy $S(\rho_{MD})$ is shown using black solid dots, while $S(\tilde{\rho}_{md})$ is shown using red solid dots. Both entropies show a plateau for $k_z$ values consistent with the one observed in Figure 3. Note that the topological states used correspond to the upper branch of energies. b) This panel shows  $|S(\tilde{\rho}_{MD}) - S(\rho_{MD})|$ using green solid dots, versus $k_z$. We calculated the modulus of the difference between the entropies with the data contained in panel a). The number next to each dot is the fidelity between both sets of matrices,  $\mathcal{F}(\rho_{MD}(k_z, L),\tilde{\rho}_{MD}(k_z, L))$. }
 \label{fig:predicted-MD}
\end{figure}

In Reference \cite{Ahmed2022}, the suggested number of measurements employed in Eq. 21 is $6^{n_q}$, where $n_q$ is the number of "qubits" over which the quantum process acts. Allocating $6^8$ vectors, each one of $2^8$ components, as the method requires, becomes impractical. In our case, it is sufficient to consider only the measurements associated with the eigenvectors of the operator given by 
\begin{equation}
\bigotimes_{i=1}^8 \sigma_x^i. 
\end{equation}

The  operator above has $2^8$ eigenvectors, a more manageable number than $6^8$. This adaptation is the only one needed to run the algorithm since the starting and predicted matrices have the same size, rendering all the Kraus operators square matrices. 

The behavior of the cost function for this case is similar to the previous one see Figure~7. The number of steps necessary to reach the required tolerance (0.1) is higher, typically around seven hundred, and $n_k=30$.  

Figure~\ref{fig:predicted-MD} shows in panel a) both entropies, $S(\rho_{MD})$ and $S(\tilde{\rho}_{MD})$, for several values of $k_z$, while panel b) shows the modulus of their difference and the fidelity between both matrices. Note that all the fidelities are better than $0.995$, and the modulus of the differences are all lower than $0.05$, which results in relative errors of less than $2\%$.

\section{Discussion and conclusions}

When dealing with one-particle wave functions or spinors, it is possible to trace out over only one of the coordinates and obtain a coordinate-dependent RDM, which leads to entropies (or information-like quantities) that detect transitions in quantum states of one-electron systems \cite{Garagiola2018,Giovenale2022a}.

In the case of a many-particle system, there are more possibilities since it is possible to trace over all the coordinates of a subset of particles (entanglement between particles) \cite{Zozulya2007,Pont2015,Haque2007} or subsets of coordinates (real-space entanglement) \cite{Sterdyniak2012,Rodriguez2009}.

Our calculation of the topological entropy exploits the fact that the eigenvectors of the $\mathbf{k}\cdot\mathbf{p}$ Hamiltonian is a spinor with four components, $|\psi\rangle$. Using the spinor, we obtain a pure density matrix $|\psi \rangle \langle \psi|$,  a  coordinate-dependent $4\times 4$  matrix. Tracing out a spatial region, as in Equation 9, leads to a coordinate-independent $4\times 4$ reduced density matrix whose eigenvalues are easy to calculate. 

Interestingly, for the topological states constructed as the spinors in Eq. 11, all the entropies that enter into the calculus of the topological entropy are constants independent of $k_z$ and $L$ except near $k_z=0$, that is, $S(\rho_{sp})=const.$, where $sp$ is any one of the triangular sectors $A, B$ or $C$, or a combination of them.  Because all these entropies are constant, it is enough to look for a quantum process such that $\rho_{ABC}=\mathcal{E}(\rho_p)$ to test the method.  

A caveat about the "constant" value of the topological entropy is in order since the statement is accurate as long as the radius of the triangular sectors $R_c$ is not too close to the radius of the nanowire, $R$, or becomes too small. 

The results in Sections 3, 4, and V correspond to a nanowire 120 nanometers in diameter and a radius $R_c=15$ nanometers. We also ran numerous numerical tests for other diameters and values of $R_c$, and the results were qualitatively the same. 

The entropy and entanglement spectrum obtained with the mode-dependent RDM are practical tools to distinguish topological from non-topological states, and we intend to test them in other contexts beyond the topological states in $\mbox{Bi}_2 \mbox{Se}_3$ nanowires. 

That the mode-dependent RDM could result from a quantum process makes us think that it is possible to define operatively an equivalent RDM, whose entropy allows us to distinguish between topological and non-topological states without resorting to a particular basis set of functions.

\acknowledgements

The authors acknowledge partial financial  support from
CONICET (PIP 11220210100787CO). and SECYT-UNC.

\appendix

\section{Variational Method} \label{ap:variational-method}

The simpler Rayleigh-Ritz variational method requires a set of appropriate basis functions that we denote as $f_i$. The expectation value of the Hamiltonian
\begin{equation}
\langle f|H|f \rangle,
\end{equation}
is obtained  using a test function given by
\begin{equation}
f=\sum_{i=1}^N c_i f_i ,
\end{equation}
and the minimization is performed over the values of the $c_i$ coefficients. This procedure results in an algebraic problem
\begin{equation}
\left[H\right] \kappa_j = E^v_j \kappa_j,
\end{equation}
where $\left[ H\right]$ is an $N\times N$ matrix whose entries are given by 
\begin{equation}
\left[ H\right]_{ij} = \left\langle f_i|H|f_j\right\rangle , 
\end{equation}
$\kappa_j$ is a vector that contains the $c$ coefficients, and the $E^v_j$ are the variational eigenvalues. $N$ is the number of functions in the basis set, and the algebraic problem has $N$ eigenvalues and their corresponding eigenvectors. 

For dealing with a  $\mathbf{k}\cdot\mathbf{p}$ Hamiltonian, like the one in Equation \eqref{eq:Hamiltonian}, the Rayleigh-Ritz method requires some adaptations. Instead of a test function, we need a test spinor to calculate the Hamiltonian expectation value. In particular, we employ the following,

\begin{widetext}

\begin{equation}
|\psi_{L,k_{z}}\rangle =\sum_{n}\left(\!\!\!\begin{array}{c}{{b_{L,k_{z},n,\uparrow}A_{L,n}J_{L}\left(\alpha_{n}^{L}\rho /R\right)e^{i L\varphi}}}\\ {{c_{L,k_{z},n,\uparrow}A_{L,n}J_{L}\left(\alpha_{n}^{L}\rho /R \right)e^{i L\varphi}}}\\ {{b_{L,k_{z,n,\downarrow}}A_{L+1,n}J_{L+1,n}\left(\alpha_{n}^{L+1}\rho/R\right)e^{i(L+1)\varphi}}}\\ {{c_{L,k_{z,n,n,\downarrow}}A_{L+1,n}J_{L+1,n}\left(\alpha_{n}^{L+1}\rho/R\right)e^{i(L+1)\varphi}}}\end{array}\!\!\!\right)e^{i k_{z}.z},
\end{equation}
where $J_L$ is the Bessel function with index $L$, $\alpha_n^L$ is its $n-th$ root, $L$ is an integer number, $R$ is the radius of the cylinder, $b_{L,k_{z,n,\uparrow}},c_{L,k_{z,n,\uparrow}},b_{L+1,k_{z,n,\downarrow}},c_{L+1,k_{z,n,\downarrow}}$ are the coefficients of the expansion (i.e., the linear variational parameters), and 
\begin{equation}
A_{L,n} = \frac{1}{\sqrt{\pi}RJ_{L+1}(\alpha_{n}^{L})}, 
\end{equation}
is a normalization constant. 

With the provisos mentioned in the paragraph above, the resulting algebraic problem for the $\mathbf{k}\cdot\mathbf{p}$ Hamiltonian has $4N$ eigenvalues and their corresponding eigenvectors. 

Finally, to calculate the matrix elements necessary for the Rayleigh-Ritz method, we write the operators $k_{\parallel}$ and $k_{\pm}$ as differential operators in cylindrical coordinates using that in Cartesian ones 
\begin{align}
    &k_{x}=-i\partial_{x}\\
    &k_{y}=-i\partial_{y}\\
    &k_{z}=k_{z}.
\end{align}
The change to cylindrical coordinates leads to
\begin{align}
    &\partial_{x}=cos(\varphi)\partial_{\rho}-\frac{sin(\varphi)}{\rho}\partial_{\varphi},\\
    &\partial_{y}=sin(\varphi)\partial_{\rho}-\frac{cos(\varphi)}{\rho}\partial_{\varphi},\\
    &\partial^{2}_{x}=cos^{2}(\varphi)\partial^{2}_{\rho}+\frac{sin^{2}(\varphi)}{\rho}\partial_{\rho}+\frac{sin(2\varphi)}{\rho^{2}}\partial_{\varphi}+\frac{sin^{2}(\varphi)}{\rho^{2}}\partial^{2}_{\varphi}\\
    &\partial^{2}_{y}=sin^{2}(\varphi)\partial^{2}_{\rho}+\frac{cos^{2}(\varphi)}{\rho}\partial_{\rho}-\frac{sin(2\varphi)}{\rho^{2}}\partial_{\varphi}+\frac{cos^{2}(\varphi)}{\rho^{2}}\partial^{2}_{\varphi},
\end{align}
which results in 
\begin{align}
    &k_{\pm}=k_{x}\pm i k_{y}=e^{\pm i \varphi} \left(-i\left(\partial_{\rho } \pm \frac{i}{\rho}\partial_{\varphi}\right)\right),\\
    &k_{\parallel}=k_{x}^{2}+k_{y}^{2}=-\left(\partial^{2}_{\rho}+\frac{1}{\rho}\partial_{\rho}+\frac{1}{\rho ^{2}}\partial^{2}_{\varphi}\right).
\end{align}

\end{widetext}


\end{document}